\documentclass{article}%
\usepackage{amssymb}
\usepackage{amsmath}
\usepackage{amsfonts}
\usepackage{mitpress}
\usepackage{graphicx}%
\providecommand{\U}[1]{\protect\rule{.1in}{.1in}}

\newdimen\dummy
\dummy=\oddsidemargin
\begin{document}

\title{\textsc{Massive Particles from Massless Spinors}}
\author{Marcus S. Cohen\\Department of Mathematical Sciences\\New Mexico State University\\Las Cruces, New Mexico\\marcus@nmsu.edu}
\maketitle

\begin{quote}
"Hark ye...the little lower layer. All visible objects, man, are but as
pasteboard masks... If man would strike, strike through the mask!" (Herman
Melville; Moby Dick).
\end{quote}

\noindent\textbf{Abstract:} Spinors are \textit{lightlike}$.$ How do they
combine to make \textit{massive }particles? We visit the zoo of\textit{\ }%
Lagrangian \textit{singularities, }or\textit{\ caustics,} in spacetime
projections from \textit{spin space- }the phase space of lightlike, \textbf{8-
s}pinor flows. We find that the species living there are the
\textit{elementary\ particles. }{\LARGE \ }Codimension $J=(1,2,3,4)$\ phase
singularities - vortex lines, sheets, tubes, and knots, are classified by the
\textit{Coxeter groups} generated by multiplicity $s$ \textit{reflections}:
"mass scatterings" off the \textit{vacuum spinors}, that keep chiral pairs of
matter envelopes confined to a timelike world tube, endowing a bispinor
particle with \textit{mass}. Using the volume in spin space as the action, the
particle masses emerge in terms of the \textit{multiplicities, s:} the number
of \textit{null zigzags }needed to close a \textit{cycle }of mass
scatterings.\textit{\ }These mass values (calculated to lowest order in the
vacuum intensity) are within a few percent\textit{\ }of the\textit{\ observed
masses }for the leptons\textit{\ }$(J=1)$ and hadrons $(J=3)$ ;\textit{\ \ }%
but are up to 25 percent off for the mesons $(J=2).$

\section{From Spin Space to spacetime}

\bigskip\textit{Spinors} live in the root space of vector and tensor fields,
much as complex numbers live in the root space of\ real polynomials.

Left ($\ell\in L$) or Right ($r\in R$) \emph{chirality }spinors have $L$ and
$R$-handed $su\left(  2\right)  $ twists with \emph{spatial }translation: they
are \emph{Clifford} $\left(  C\right)  $-analytic and conjugate $C$-analytic,
respectively [BdS]. \textit{Complexified} spinors [M.C.1] have an additional
counter-clockwise or clockwise $u\left(  1\right)  $ phase advance with
\textit{cosmic} \textit{time}$,T:$ the log of the distance back to the big
bang, in units of the compactification radius, $a_{\#}:$

\begin{center}
$T=a_{\#}\ln\gamma$; $\ \gamma\left(  t\right)  =\frac{alt)}{a_{\#}}$ .
\end{center}

Here $a(t)$\ is the scale factor of our spatial hypersurface, $S_{3}[a(t)],$as
a function of $Minkowsky$ time, or \textit{arctime, }$t:$ $\left\vert
\bigtriangleup\mathbf{x}\right\vert =\pm\bigtriangleup T=\bigtriangleup t,$
the distance traveled by a photon.. Arctime and cosmic time combine to make
\textit{complex time}, $z^{0}\equiv t+iT.$ A spinor is either \emph{analytic}
$\left(  +\right)  $ or \emph{conjugate analytic} $\left(  -\right)  $ in
complex time; the sign, $Sgn[\partial_{T}\theta^{o}]$ is its $charge$
[MC.1][M.C.2]. Complex conjugation, $\left(  z^{0}\right)  ^{\ast}\equiv
t-iT$, \ gives charge reversal, C.

There are \textbf{8} Complex Clifford-algebra representations of the
spin-isometry group of a 4 -manifold: the Einstein group, \textbf{E}:
4\textbf{\ }column spinors $\psi_{I}=\{l_{+},r_{+},l_{-},r_{-}\}$, and
4\textbf{\ (}provisionally independent) row spinors, $\psi^{I}=\{l^{+}%
,r^{+},l^{-},r^{-}\}.PT$ reversal$,$ $PT:q\rightleftarrows\overline{q},$
$z\leftrightarrows z^{\ast}$ gives an equivalent representation, there are 2
spinors in each class:%

\begin{equation}%
\begin{tabular}
[c]{lll}
& $q$ & $\overline{q}$\\\cline{2-3}%
$z_{0}$ & \multicolumn{1}{|l}{$\ell^{+}\text{, }r_{-}$} &
\multicolumn{1}{|l|}{$r^{+}\text{, }\ell_{-}$}\\\cline{2-3}%
$z_{0}^{\ast}$ & \multicolumn{1}{|l}{$\ell^{-}\text{, }r_{+}$} &
\multicolumn{1}{|l|}{$r^{-}\text{, }\ell_{+}$}\\\cline{2-3}%
\end{tabular}
\end{equation}

where $\overline{q}_{a}\equiv(q_{0},-q_{1},-q_{2},-q_{3})$ (notation as in
[M.C.1], [M.C.2], [M.C.3]).

$PT$-conjugate spinors have the same \textit{helicities}:%

\[%
\begin{array}
[c]{cc}%
\left(  \ell_{+},r_{-}\right)  \subset\text{ }\circlearrowright &
\text{(\emph{left} helicity)}\\
\left(  \ell_{-},r_{+}\right)  \subset\text{ }\circlearrowleft &
\text{(\emph{right} helicity)}%
\end{array}
\text{.}%
\]

The $L$ and $R$-chirality spinors in an electron, $e_{-}\equiv(l_{-}\oplus
r_{-}),$ have \emph{opposite helicities}, but the \emph{same spins.} They are
thus \emph{counterpropagating}. This explains their opposite boost dependence.
For central bispinors, one propagates \emph{outward}, in the direction of
cosmic expansion, $\left\vert \Delta\mathbf{x}\right\vert =\Delta T>O,$ and
the other \emph{-}$\left\vert \Delta x\right\vert =\Delta T<O$ propagates
\emph{inward}. Similarly, we identify direct sums of \emph{copropagating}
spinors of opposite chirality, but the \emph{same helicities}, as
\emph{neutrinos}$;$%
\[
\nu=(l_{-}\oplus r_{+});e_{-}=\left(  l_{-}\oplus r_{-}\right)  .
\]

Tensor products of 2 opposite-chirality spinors make null spin vectors:
Photons. The \textit{null tetrads }are\textit{\ }"vacuum photons", of helicity
$\pm$1:%

\begin{equation}
q_{\upharpoonright}\equiv l_{+}\otimes r^{-};q_{\downarrow}\equiv l_{-}\otimes
r,\ q_{+}\equiv l_{+}\otimes r^{+};q_{-}\equiv l_{-}\otimes r^{-},
\end{equation}

Spin-1 sums of these make the \textit{Clifford tetrads }of a moving Clifford
algebra (C) frame:%

\begin{equation}%
\begin{array}
[c]{c}%
q_{a}\equiv\ell\otimes r\in CT\mathbb{M}:\ \ \ \ q_{o}\equiv
(q_{\upharpoonright}-q_{\downarrow)};\ \ q_{1}\equiv(q_{+}+q_{-});\\
q_{2}\equiv-i(q_{+}-q_{-});\ \ q_{3}\equiv(q_{\upharpoonright}%
+q_{\downharpoonright}).
\end{array}
\end{equation}

The $q_{a}$ are identified with the basis vectors, $e_{\alpha},$ of a
spacetime frame via the\textit{\ spin map, }$S,$ whose pullback is
the\textit{\ Dirac operator, }$D=S^{\ast}.$\textit{\ }On\textit{\ compactified
Minkowsky space}, $\mathbb{M}_{\#}=S_{1}\times S_{3}(a_{\#}),$%

\begin{equation}
S=(\frac{i}{2a_{\#}})q^{\alpha}\partial_{\alpha}e_{\alpha}\rightarrow\left(
2a_{\#}\right)  -1q_{\alpha};\quad D=\ \ S^{\ast}:\left(  2a_{\#}\right)
e^{\beta}\hookleftarrow q^{\beta}%
\end{equation}

More generally, in moving frames in spin space and spacetime, the spin map reads%

\begin{equation}%
\begin{array}
[c]{c}%
S(x)\equiv\lbrack\partial_{\alpha}\zeta^{\beta}](x)q_{\beta}(x)e^{\alpha
}(x):e_{\alpha}(x)\rightarrow\left[  \partial_{\alpha}\zeta^{\beta}\right]
q_{\beta}(x);\\
\lbrack\partial_{\alpha}\zeta^{\beta}](x)\equiv\left[  d\zeta\right]
(x),\text{ where }\mathbf{d}\equiv e^{\alpha}d_{\alpha}%
\end{array}
.
\end{equation}

is the generalized exterior differential operator. The Jacobean determinant,
$\left\vert d\zeta\right\vert (x),$ is the $4$- volume expansion factor;
$\left\vert d\zeta\right\vert =0$ at a singular (critical) point, $x_{c}$ (we
will suppress the generic point, $x$ $=(T,x^{1},x^{2},x^{3})$
below).\textit{\ }

General covariance says that the equations of motion must be covariant, and
their action integrals\ invariant, under coordinated spin isometries of
spacetime (external) and spin space (internal) frames.\textit{\ \ }

Covariance is \textit{automatic }if our spacetime is a horizontal local
section of an 8\textbf{- }\textit{spinor bundle, }\textbf{8;} a\textit{\ }%
different section for an observer in a different frame.\textbf{\ }We treat the
spinors here as the \textit{real, physical }objects, and spacetime vector and
tensor fields as their preimages under the\textit{\ }spin\textit{\ }%
map\textit{, }$S.$ We call this the

{\LARGE Spin principle,(P1). }The 8-spinor bundle, \textbf{8} is the real
physical object. Spacetime, geometry, gauge, and matter fields, along with
their interactions, all emerge as projections of $J$ chiral pairs of matter
envelopes, $(\psi^{I},\mathbf{d}\psi_{I},\mathbf{d}\psi,^{I}\psi_{I}),$
selected from the $\mathbf{4}$ left and $\mathbf{4}$ right chirality
spinors\textit{\ }and their differentials. Their coupling constants are
products of the remaining $(4-J)$ \textit{vacuum }pairs, chosen from the
4\textbf{\ }column spinors, $\psi_{I}=\{r_{-},l_{-},r_{+},l_{+}\}$, and the
$\left(  4-J\right)  $ \textit{provisionally independent} row spinors
$\psi^{I}\equiv\{l^{+},r^{+},l^{-},r^{-}\}.$

\qquad

\qquad

It takes \textbf{4} spinors to make the pseudoscalar "inner product,"%

\begin{equation}
\psi^{I}[iq_{2}(x)]\psi_{I}\equiv\psi^{I}i[\hat{l}(x)\otimes_{2}\hat
{r}(x)]\psi_{I},
\end{equation}

using $[q_{2}(x)]$ as a matrix in the \textit{moving} spin frames. On a curved
space, there is no C scalar bilinear for like $\bar{\psi}\psi_{\lambda},$ so
no dualizing operation like $\bar{\psi}=\psi^{T}\epsilon=\left(  \psi_{2}%
-\psi_{1}\right)  $ in flat space.

It takes \textbf{4}$\ $spinors make the \emph{metric tensor} (spin 2); it
takes \textbf{8} spinors to make an \textbf{E-} invariant inner product; a
\ \textit{C scalar}, $q_{0}:$%

\begin{equation}
g_{\alpha\beta}=\frac{1}{2}[q_{\alpha}\otimes\overline{q}_{\beta}\oplus
q_{\beta}\otimes\overline{q_{\alpha}}];\ \ q^{\alpha}g_{\alpha\beta}q^{\beta}.
\end{equation}

Each spinor has conformal weight (dimension) $\frac{1}{2},$ so it takes
\textbf{4 }spinors and \textbf{4} spinor differentials to make the simplest
\textbf{\ E}-invariant Lagrangian 4 form, with \textit{no coupling constants}:
\ the \textbf{8}\textit{-spinor} \textit{factorization of the bi-invariant}
\textit{Maurer-Cartan (M.C.) 4 form, }%

\begin{equation}
\mathcal{L}_{g}=\left[  (\psi^{1}\mathbf{d}\psi_{1})\wedge(\mathbf{d}\psi
^{2}\psi_{2})\wedge(\psi^{3}\mathbf{d}\psi_{3})\wedge(\mathbf{d}\psi^{4}%
\psi_{4})\right]  ^{0},
\end{equation}

where $\left[  C\wedge^{4}\right]  $ means "the scalar part" of a C-valued 4 form.

Here $\mathbf{d}\equiv e^{\alpha}\left(  x\right)  \partial_{\alpha}\left(
x\right)  $ is the generalized (possibly path-dependent) exterior differential
operator. It includes the derivatives of the \emph{moving spin frames,} and
reduces to the covariant derivative, coordinate wise:%

\begin{equation}%
\begin{array}
[c]{c}%
\mathbf{d}\psi_{I}\equiv\mathbf{d}\left(  \ell^{I\mathbf{\ \ }}{}\psi
_{I}\right)  =l^{I}\mathbf{d}\psi_{I}+\mathbf{d}l^{I}\psi_{I}=l^{I}\left(
\partial_{\alpha}+\Omega_{\alpha}\right)  \psi_{I}e^{\alpha}\\
\equiv l^{I}\nabla_{a}\psi_{I}e^{\alpha}.
\end{array}
\end{equation}

The action integral is the\textit{\ volume }of the state, $\Psi\equiv(\psi
_{I},\mathbf{d}\psi_{I}),(\psi^{I},\mathbf{d}\psi^{I})\subset T^{\ast}\Sigma,$
in\textit{\ spin space}: the phase space of \textbf{8}-spinor flows\textit{.}
In\textit{\ }complex coordinates on $T^{\ast}\Sigma,$ $\ $%

\begin{equation}
S_{g}=\frac{1}{2i}\int_{M_{\#}}\left[  (\psi^{I}-i\mathbf{d}\psi^{I}%
)\wedge(\psi_{I}+\mathbf{d}\psi_{I})\right]  ^{4}\rightarrow PT\rightarrow
\int_{M_{\#}}\left[  (\psi^{I\mathbf{\ }}\mathbf{d}\psi_{I})\right]  ^{4},
\end{equation}

in either the $PT$-symmetric $(PT_{s})$ case, or the $PT$-antisymmetric
$\left(  PT_{a}\right)  $ case, $\psi^{I}\mathbf{d}\psi_{I}=\pm\mathbf{d}%
\psi^{I}\psi_{I}$ Stationarizing $S_{g}$ gives a minimal 4-surface, $\Psi\in
T^{\ast}\Sigma.$

In order to be localized inside a compact world tube, $B_{4}$ the matter
spinors must match the vacuum distribution on its boundary, $\gamma_{3}%
\equiv\partial B_{4}$ \ [Taubs], [Uhlenbeck]. In the
regular,\textit{\ geometrical-optics} regime outside, $S_{g}$ yields the
proper effective actions for electroweak (PTa, or charge-separated) and
gravitostrong (PTs , or neutral) fields, $\mathbf{dd}\psi=\mathbf{\kappa\psi
.}$ Here it agrees with Witten's "\emph{Weiss-Zumino} 4 form," $Tr(g^{-1}%
dg)^{\wedge4}g$ , whose action is quantized over the boundary, $\gamma_{4}%
\sim\partial B_{5},$ of a 5-manifold [Witten1], [Witten2]. Here, we find
$B_{5}\sim\mathbb{C}\times\mathbb{S}_{3}$ embedded in the position-world
velocity phase space, $z^{\alpha}=x^{\alpha}+y^{\alpha},$ with complex time
coordinate, $z^{o}=t+iT$.

\bigskip Geometrical optics break down on \textit{boundary caustics,}
$\gamma_{4-J}$ $\in$ $\partial B_{5-J}$, where, the spin map $S:$
$TM\rightarrow T\Sigma$ becomes \textit{singular, }and acquires a
$J-$dimensional kernel. The domains these caustics enclose are
\textit{branched covers, }$B_{4-J}$\textit{\ }$\equiv\ast D^{J},$ with $J$
extra\textit{\ }bispinor sheets in spin space over each spacetime point $x\in
B_{4-J}.$ These accommodate the wave functions of $J-$bispinor particles.

Caustics arise in optics, hydrodynamics, chemical reactions, acoustics, etc.
as loci of partial focusing, or shock fronts [Arnold]. Joe Keller, Alan Newell
[Newell], and others have used a powerful tool to look inside these apparent
singularities: \textit{singular perturbation theory} or multiscaling; defining
a short spacetime scale inside the shock, and matching the inner solution to
the outer one on the shock boundaries. We apply it to give a system of coupled
envelope-modulation equations [Newell] to nonlinear waves in the 8 spinor
medium: the \textit{spinfluid, }and find\textit{\ }that their caustics are the
\textit{elementary particles}. We outline the results below; details of the
calculations appear in Part III [M.C. 4].

\section{Singularities and Stratification}

\bigskip In the g\textit{eometrical-optics (g.o.) }regime, $D^{0},$
\textit{regular} phase flows are created by nonsingular active-local (perhaps,
path-dependent) Einstein transformations, $\left(  L(x),R(x)\right)  $\ $\in
E_{A}$, acting on the \textit{vacuum spinors}, $\left(  \hat{\ell},\hat
{r}\right)  ,$ written as $GL(2,C)$ matrices column wise and row wise respectively:%

\begin{equation}%
\begin{array}
[c]{c}%
\ell\left(  x\right)  =\hat{\ell}\exp\left[  \frac{i}{2}\zeta_{L}^{\alpha
}q_{\alpha}\right]  \equiv\hat{\ell}L\left(  x\right)  ;\\
r\left(  x\right)  =\exp\left[  \frac{i}{2}\zeta_{R}^{\alpha}\overline
{q}_{\alpha}\right]  \hat{r}\equiv R\left(  x\right)  \hat{r}\text{.}%
\end{array}
\end{equation}

In the $PTa$ case, $R\left(  x\right)  =L^{-1}\left(  x\right)  ,$ multiplying
a spinor by the differential of the PT opposed spinor gives effective
\textit{spin connections}$:$ C- algebra-valued 1 forms; or \textit{vector
potentials }[Keller]:%

\begin{equation}%
\begin{array}
[c]{c}%
\Omega_{L}\equiv\ell^{-1}\mathbf{d}\ell\left(  x\right)  =\mathbf{d}\zeta
_{L}=\left[  \partial_{\alpha}\zeta_{L}^{a}\right]  (x)q_{a}e^{\alpha}%
\text{,}\\
\Omega^{R}\equiv\left(  \mathbf{d}r\right)  r^{-1}\left(  x\right)
=\mathbf{d}\zeta_{R}=\left[  \partial_{\alpha}\zeta_{R}^{a}\right]
(x)\overline{q}_{a}e^{\alpha}\text{.}\\
\end{array}
\end{equation}

However, even for a regular initial distribution of spinor fields,
codimension- $J=\left(  1,2,3,4\right)  $\emph{\ singular Lagrangian }%
$\equiv\ast\gamma\left(  4-J\right)  $, will form, shift, merge, annihilate,
and recombine, like the projections of folds in a sheet to the bed. In
addition to the \textit{regular stratum, }$\gamma^{o},$where the projection
from the Lagrangian submanifold of spin space solutions to the position-world
velocity phase space,%

\[
\pi:(\psi_{I}+i\mathbf{d}\psi_{I})\in%
\mathbb{C}
T^{\ast}\Sigma\rightarrow x^{\alpha}+iy^{\alpha}\in%
\mathbb{C}
T^{\ast}M,
\]
is 1 to 1, there will be codimension-$J=(1,2.3,4)$ \textit{branched covers,}
$D^{J},$ where $\pi$ is $\ J+1$ to $1.$ Like the crisscrossing rays inside a
kaleidoscope there are $J+1$ world-velocity sheets, $y^{\alpha},$ over each
spacetime point, $\ x^{\alpha}\in B_{4-J}$, in the supports,\textit{\ }%
$B_{4-J},$ of $J-$bispinor particles.$.$ Each support is bounded by\ loci of
partial focusing,\emph{\ }boundary caustics\emph{,}$\gamma_{4-J}$%
\emph{\ }$\subset\partial B_{5-J}$ : \ folds, cusps, tucks, swallow-tails and
knots, where spin rays $\psi^{I}\mathbf{d}\psi_{I}$ = $\mathbf{d}\zeta_{I}$
branch or converge [ref Arnold]. Each $(4-J)$ \textit{brane,} $B_{4-J},$
carries a $J-$form matter current, $\ast J,$\textit{dual} to the Clifford
volume element contributed by the $(4-J)$\textit{\ vacuum }pairs\textit{.} We
call this complex of$\ $branes and currents the Spin (4,$%
\mathbb{C}
$) \textit{complex}, or \textit{spinfoam}.\qquad\qquad\qquad\qquad\qquad
\qquad\qquad

A 3- dimensional example is a foam of soap bubbles, with the regular stratum,
$B_{3}=\ast D^{o}$ (the volumes), and singular strata, $\gamma_{2}%
\subset\partial B_{3},\gamma_{1}\subset\partial B_{2},\gamma_{0}%
\subset\partial B_{1}:$ the surfaces, edges, and vertices. Each stratum,
$D^{J},$carries a $\ J-$ form current: density in volumes $\gamma_{3}$
pressure on surfaces $\gamma_{2},$ tension in line segments $\gamma_{1}$, and
force on nodes $\gamma_{0}.$

A codimension- J $bifurcation$ occurs at the critical point, $x_{c}\in
\gamma_{4-J},$where the rank of the Jacobian matrix, $\ [\mathbf{d\zeta
}\boldsymbol{]}(x_{c})\boldsymbol{\equiv}$ $[\partial_{\alpha}\zeta_{\beta
}](x_{c}),$ drops by J. Here, $J+1$ phase differentials become linearly
dependent, to span only a $\left(  4-J\right)  $-dimensional subspace. If the
Hessian, $[\mathbf{d}^{2}\zeta](x_{c})$, is singular there too, $\left\vert
\partial_{\alpha}\partial_{\beta}\zeta\right\vert (x_{c})=0$,\textbf{\ }%
$x_{c}$ is a\textit{\ degenerate critical point: }a\textit{\ }caustic, where
rays $\mathbf{d\zeta}_{I}$\ merge or split, and there is a change in the
topology of the orbits..\qquad

This is \textit{dynamical} \emph{symmetry breaking. }One tool to detect it is the

\qquad\qquad\qquad\qquad\qquad\qquad\qquad\qquad\qquad\qquad\qquad\qquad\qquad

$\mathbf{Equivariant\ Branching\ lemma}$ (Michel's "theorem"): If the isotropy
subgroup, $H\subset E,$ that fixes a solution $\Psi_{{\Large c}}$ contains
just a single copy of the \emph{identity} representation, then $\Psi_{C}$ is a
possible direction for dynamical-symmetry breaking ref. [Sattinger].

Some corollaries are

\begin{enumerate}
\item the branched covers and boundary caustics \emph{stratify} the base
space, $\mathbb{M}$, into orbits of $E$-group actions into \emph{isotropy
subgroups,} $H:$%
\[
\mathbb{M}=\bigcup_{J=0}^{4}B_{4-J}\oplus\gamma_{4-J}.
\]

\item Generically, as you cross a \textit{boundary\ caustic} $\gamma
_{3-J}\equiv\partial B_{4-J},$ where \
$\vert$%
$dd\zeta|=0,\ker d\zeta$ picks up generators one at a time

\item The boundary of each stratum consists of singular loci belonging to the
next higher stratum,\emph{\ }except where two caustics intersect. Here, their
co-dimensions add:%
\[
\gamma_{4-J}\cap\gamma_{4-K}=\gamma_{4-M}:M=J+K\text{.}%
\]

\end{enumerate}

Gluing conditions for splicing a compact "bubble", $\Omega^{J}\equiv(\psi
^{I}d\psi_{I})^{J}$\ of $J$ matter-spinor pairs into the vacuum distribution
give constraints on\ their integrals. As the "neck" of the J-tube
$\gamma_{4-J}$ joining the matter bubble and the vacuum background contracts
to (or expands from) a single point, the matter wave functions must
\textit{match} the vacuum spinors there. This demands integral periods for J-
form matter currents over compactified $J$ cycles: quantized
\textit{topological charges}. [Taubs] , [Uhlenbeck].

The \textbf{Fibration Theorem}\textit{\ }[Milnor] guarantees a complete set of
\ $(4-J)$ \ \textit{parallelizable}\ fiber coordinates bridging the gaps
between codimension-$J$ singular loci: the integral curves of \ the
$\emph{vacuum\ spin\ forms,\ }\widehat{\Omega}^{4-J}$ (Table I).

\newpage

\begin{center}
{\Large Table I: the\ vacuum\ spin forms,}

Assuming the vacuum spinors all have the same amplitude, $k^{\frac{1}{2}},$%

\[%
\begin{array}
[c]{c}%
\hat{\Omega}=\pm\left(  \frac{ik}{2a}\right)  q_{\alpha}e^{\alpha}\\
\hat{\Omega}^{2}=\left(  \frac{ik}{2a_{\#}}\right)  ^{2}q_{\ell}\left[
\epsilon_{jk}^{\;\ell}e^{j}\wedge e^{k}\pm e^{0}\wedge e^{\ell}\right] \\
\hat{\Omega}^{3}=\pm\left(  \frac{ik}{2a_{\#}}\right)  ^{3}q_{\ell}%
\epsilon_{jk}^{\;\ell}e^{j}\wedge e^{k}\wedge e^{0}\pm i\epsilon_{jk\ell}%
q_{0}e^{j}\wedge e^{k}\wedge e^{\ell}\\
\hat{\Omega}^{4}=\left(  \frac{ik}{2a_{\#}}\right)  ^{4}q_{0}\left[
\epsilon_{\alpha\beta\gamma\delta}e^{\alpha}\wedge e^{\beta}\wedge e^{\gamma
}\wedge e^{\delta}\right]  =\frac{3}{2}\left(  \frac{k^{4}}{a_{\#}^{4}%
}\right)  \mathbf{d}^{4}V,
\end{array}
\]

\bigskip
\end{center}

The constraint that the Lagrangian density must be a C scalar assures that
only the parts of $\hat{\Omega}^{4-J}$ both \textit{Clifford }and
\textit{Hodge} \emph{dual} to the matter forms, $\tilde{\Omega}^{J}%
\equiv\left(  \psi^{I}\mathbf{d}\psi_{I}\right)  ^{J},$ to contribute to the
action. These make the Clifford line, surface, and volume elements that
multiply $\tilde{\Omega}^{J}\ $to fill out the\textit{\ }\textbf{E-}invariant
(C-scalar) $4$\emph{\ -}volume element,

\begin{center}%
\[
\left\vert (\mathbf{d}\zeta)^{4}\right\vert \sigma_{o}e^{0}\wedge e^{1}\wedge
e^{2}\wedge e^{3}:\ \thicksim\gamma^{4}(dx)^{4}.
\]

\end{center}

Any C-dual contribution to $S_{g}$ must therefore be Hodge dual, as well,
effectively quantizing $\tilde{\Omega}^{J}$ against \emph{dual}
(perpendicular) cycles, $\gamma_{4-J}$ , as well as over cycles $\gamma_{J} $
\ (e.g. \ quantization of electric, flux, $F_{or}$ , over $S_{2}(\theta,\phi)$
[M.C. 2]).

These topological charges remain constant with cosmic expansion, while the
vacuum spin forms\emph{,} $\hat{\Omega}^{4-J}$ (table I) give a factor of
$\ k^{4-J}\sim\gamma^{J-4}$ \ to the action contributed by the $D^{J}$
stratum. \ Integrating in the comoving frame, $E^{\alpha}=\gamma e^{\alpha} $
results in a net action polynomial in the scale factor, $\gamma:$ the
\textit{effective potential},%

\begin{equation}%
\begin{array}
[c]{c}%
V(\mathbf{n,\gamma)}=\int_{D_{0}}\hat{\Omega}^{4}+\gamma\int_{D_{1}}%
\hat{\Omega}^{3}\wedge\left(  \psi^{I}\mathbf{d}\psi_{I}\right)  +\gamma
^{2}\int_{D_{2}}\hat{\Omega}^{2}\wedge\left(  \psi^{I}\mathbf{d}\psi
_{I}\right)  ^{2}\\
+\gamma^{3}\int_{D_{3}}\widehat{\Omega}\wedge\left(  \psi^{I}\mathbf{d}%
\psi_{I}\right)  ^{3}+\gamma^{4}\int_{D_{4}}\left(  \psi^{I}\mathbf{d}\psi
_{I}\right)  ^{4}=16\pi^{3}\left[  n_{0}+n_{1}\gamma+n_{2}\gamma^{2}%
+n_{3}\gamma^{3}+n_{4}\gamma^{4}\right]  ,
\end{array}
\end{equation}

where $n_{J}$ is the population of the $Jth$ stratum [M.C..3] .

The polynomial $V(\mathbf{n,}\gamma)$ can \textit{mimic the effect of the
Higgs field }by mixing positive-definite quadratic couplings in $\gamma^{2}$
with \ negative-definite quartic ones in $-$ $\gamma^{4},$ to create a
"Mexican hat" potential. But, unlike standard Q. F. T., the lepton, meson,
hadron and atomic masses appear in a\textit{\ }4-term sequence, at
$O(\gamma,\gamma^{2},\gamma^{3,}\gamma_{4}),$ respectively.

The $\hat{\Omega}^{3}$ term contributes the 3 -volume element in \textit{spin
space }to\ the Noether charge under complex-time $(z^{0}\equiv t+iT)$
translation, which includes the Jacobean determinant of the $3$-space block of
spin map, $S:$%

\[
|\left(  \mathbf{d\zeta}\right)  |^{3}\thicksim s^{3}e^{1}\wedge e^{2}\wedge
e^{3}.
\]

This gives quantization of both \textit{mass }and\textit{\ charge}:%

\begin{equation}
\int_{B_{3}}[(\partial_{t}\theta^{0})-i(\partial_{r}\theta^{0})]e^{1}\wedge
e^{2}\wedge e^{3}=M+iQ.
\end{equation}
It is the\textit{\ vacuum spinors, } hiding the C 3-volume element
$\hat{\Omega}^{3}$ that endow frequency, $\omega\equiv(d_{t}\theta^{0}),$
with\textit{\ mass:\ Mach's principle} in action.

Gluing principle P2. says that the matter spinors localized inside the compact
world tube $B_{4}$ the matter spinors must match the vacuum distribution on
its boundary, $\gamma_{3}\equiv\partial B_{4\text{ }}.$

As you pass through a degenerate codimension- $J=(1,2,3,4)$ singularity,
$x_{c}\in$ $\gamma_{4-J}\subset\partial B_{5-J},$ both the Jacobean and the
Hessian determinants vanish, and the rank of the spin map drops by $J$:%

\begin{equation}
S\equiv\left[  \partial_{\alpha}\zeta^{\beta}\right]  \left(  x_{c}\right)
\ q_{\beta}e^{\alpha}:\left\vert \partial\zeta\right\vert \left(
x_{c}\right)  =\left\vert \partial^{2}\zeta\right\vert (x_{c})=0\Rightarrow
\ r(x_{c})=(4-J).
\end{equation}

A point inside\ $\gamma_{4-J}$ \ acquires $2J$ new preimages in the projection
$\pi:L\rightarrow M$\ \ from the Lagrangian submanifold in spin space to spacetime.

To look inside the singular loci, we use \textit{singular perturbation theory;
}what Don Cohen calls \textit{"two timing and double crossing"}. Following Joe
Keller, Alan Newell [Newell], and others, we defining a short spacetime scale,
$x=\gamma X$ inside the shock front, and match the inner solution to the outer
one on the shock boundaries. We apply it here to caustics in the
\textit{spinfluid, }which turn out to be the elementary particles. We outline
the results here; details of calculations appear in Part III [M.C.4].

First, we express each spinor field as a vacuum field, $\varphi_{I}%
\equiv\left(  \hat{\ell}_{\pm},\hat{r}_{\pm}\right)  $ of amplitude
$k^{\frac{1}{2}}\thicksim\gamma^{-\frac{1}{2}},$ plus an envelope modulation:%

\begin{align}
\ell_{I}(x,X)  &  =k^{\frac{1}{2}}\hat{\ell}_{\pm}(X)+\psi_{L}^{\pm
}(x)\ =\gamma^{-\frac{1}{2}}\hat{\ell}_{\pm}(X)+{\Huge \ }\psi_{L\pm}(x)\ ;\\
r_{\pm}\left(  x,X\right)   &  =\gamma^{-\frac{1}{2}}\left(  X\right)
+\psi_{R\pm}.\nonumber
\end{align}

In inflated regimes, like ours, $\gamma\gg1.$ In superdense regimes,
$\ \gamma$ $\ll1;$ the matter spinors are ripples riding on the vacuum: a deep
ocean of \emph{dark energy}. Since solutions are either symmetric or
antisymmetric about the critical radius, $a=a_{\#};$ $\gamma=1$, we can
consider either case, and cover both [M.C.1]. Inserting \textit{ansatz
}$\mathbf{(}17\mathbf{),}$we obtain effective Lagrangians, $\mathcal{L}^{J}%
$\emph{,} in which $(4-J)$ vacuum pairs couple $J$ matter pairs. Varying with
respect to $\hat{\ell}_{\pm}$ or $\hat{r}_{\pm}$ gives the \emph{massless
Dirac equations}. These say that the vacuum spinors are
\textit{Clifford-analytic} and conjugate-analytic respectively:%

\begin{equation}%
\begin{array}
[c]{c}%
\overline{D}\hat{l}_{\pm}\equiv q^{\alpha}\left(  \partial_{\alpha}%
+\hat{\Omega}_{a}^{R}\right)  \hat{l}_{\pm}\left(  X\right)  =O\\
D\hat{r}_{\pm}\equiv q^{\alpha}\left(  \partial_{\alpha}+\hat{\Omega}_{a}%
^{L}\right)  \hat{r}_{\pm}\left(  X\right)  =O.
\end{array}
\end{equation}

\emph{Covariantly constant} (freely-falling) solutions, $\left(
\partial_{\alpha}+\Omega_{\alpha}\right)  \left(  \hat{l}_{\pm},\hat{r}_{\pm
}\right)  =0$ define \emph{inertial spin frames}$.$ On $M_{\#}\equiv
S_{l}xS_{3}(a_{\#}),$%

\begin{equation}%
\begin{array}
[c]{cc}%
\hat{\ell}_{\pm}\left(  X\right)  =\mathbf{\hat{\ell}}_{\pm}\left(  0\right)
\exp(\frac{i}{2a_{\#}}X^{\alpha}\sigma_{\alpha}^{\pm}); & \hat{r}_{\pm}\left(
X\right)  =\hat{r}_{\pm}\left(  0\right)  \exp(\frac{i}{2a_{\#}}X^{\beta
}\overline{\sigma}_{\beta}^{\pm});\\
\hat{\Omega}_{\pm}^{L}=\frac{i}{2a_{\#}}\sigma_{\alpha}^{\pm}e^{\alpha}, &
\hat{\Omega}_{\pm}^{R}=\frac{i}{2a_{\#}}\overline{\sigma}_{\beta}^{\pm
}e^{\beta}.
\end{array}
\end{equation}

For a given scale factor, $\gamma,$ the vacuum action is extremized when the
inertial spinors span a\textit{\ hypercube} in spin space.

Neutral combinations of vacuum spinors could be called "cosmological
neutrinos", $\nu_{l}=(\widehat{l}_{+}\oplus\widehat{r}_{-});v_{r}=(\widehat
{l}_{-}\oplus\widehat{r}_{+}).$ More generally, \emph{left} and \emph{right}
\emph{chirality} moving spin frames, $\ell_{\pm}$ and $r_{\pm}$, are given
by\emph{\ path-dependent,} \emph{active-local\ }($E_{A}$)
\emph{transformations} on the inertial spinors [M.C. 1], [M.C. 2], [M,C. 3].
These vary on the cosmic scale, so $\gamma$ beats of the logic clock, $\Delta
X^{0}=\gamma,$ elapse for each beat, $\Delta x^{0}$ $=1,$ of the local clock.%

\begin{equation}%
\begin{array}
[c]{c}%
\ell\left(  X,x\right)  \equiv\hat{\ell}_{\pm}\left(  X\right)  L^{\pm}\left(
x\right)  \text{;}\\
r\left(  X,x\right)  =\hat{r}_{\pm}\left(  X\right)  \bar{R}^{\pm}\left(
x\right)  .
\end{array}
\end{equation}

At $O(\gamma),$we obtain the massive Dirac system as our coupled-envelope
equations. Dirac mass - chiral cross coupling - appears via a \emph{spin}
$\left(  4,\mathbb{C}\right)  $ \emph{resonance}; the 8-spinor analog of
4-wave mixing in nonlinear optics [M.C. 5] .

To contribute a $C$ scalar 4 form $\sigma_{0}e^{0}\wedge e^{1}\wedge
e^{2}\wedge e^{3}$ to the action integral, a chiral pair of matter spinors
must find 3 \textit{other} pairs of vacuum spinors whose product meets the
\emph{Bragg }$\emph{(solvability)}$\emph{\ conditions; the massive Dirac
equations,}

.%
\begin{equation}%
\begin{array}
[c]{c}%
\overline{D}\psi_{I}^{L}\equiv q^{\alpha}\left(  \partial_{\alpha}%
+\tilde{\Omega}_{a}^{L}\right)  \psi_{I}^{L}\left(  X\right)  =[2a_{\#}%
]^{-1}\psi_{I}^{R}\\
D\psi_{I}^{R}\equiv q^{\alpha}\left(  \partial_{\alpha}+\tilde{\Omega}_{a}%
^{R}\right)  \psi_{I}^{R}\left(  X\right)  =[2a_{\#}]^{-1}\psi_{I}^{L}.
\end{array}
\end{equation}

The electron mass-the inverse of the critical diameter, $2a_{\#},$ comes from
the product of the 3 unbroken vacuum pairs; the $\hat{\Omega}^{3}$ in Table l.
If the vacuum spinors have different amplitudes, the scalar mass term is
replaced by the term $\psi^{I}[\hat{\Omega}^{3}]_{I}^{J}\psi_{J},$ in
the\textit{\ lepton mass matrix}$.$ This is a rank- 2 tensor product of the 6
remaining vacuum spinors C dual to $(\psi^{I},\psi_{J});$ the ones needed to
make the $\ $C-scalar $(\sigma_{0})$ term, at $O(\gamma):$ $\psi^{I}%
[\hat{\Omega}^{3}]_{I}^{J}\psi_{J}\in\langle2,\widehat{\mathbf{6}}\rangle\in
L^{1}.$

\bigskip At $O(\gamma^{2}),$integration by parts gives wave equations in
$(\overline{D}D+D\overline{D})\equiv\Delta:$ Klein -Gordon (spin l or 0)
equations, sourced in the baryon -current $3$ form, $J$, quantized over 3 -cycles:%

\begin{equation}
(\Delta+\hat{\Omega}^{2})\widetilde{\Omega}=J;\quad\int_{B_{3}}J=\int_{B_{3}%
}[\partial_{o}\widehat{\theta}_{o}]\mathbf{d\zeta}^{l}\wedge\mathbf{d\zeta
}^{2}\wedge\mathbf{d\zeta}^{3}=8\pi^{2}[2a_{\#}]^{-1}B,
\end{equation}

where $B$ is the\textit{\ Baryon number. }Again, the mass term, $\hat{\Omega
}^{2},$ comes from the vacuum energy. For photons, $B=0.$

At $O(\gamma^{3}),$the principal part is a system of 3 Euler equations,
coupling each quark current, $Q_{I},$ to the 2 others through the vacuum
spinors:%
\[%
\begin{array}
[c]{cc}%
\qquad Q_{I}\in\lbrack l\otimes_{I}r]l;\ Q^{I}\in r[r\otimes^{I}l]: &
\qquad\overline{D}Q_{I}\thicksim\lbrack2a_{\#}]^{-1}T_{KI}^{J}Q_{J}Q^{K}.\\
& \qquad DQ^{I}\thicksim\lbrack2a_{\#}]^{-1}T_{J}^{IK}Q^{J}Q_{K}%
\end{array}
\]

$T$ is the rank -3 "moment of inertia" tensor, with eigenvalues $I\equiv
(\ p,q,r).$ Orbits lie on invariant tori or ellipsoids, and \textit{close for
integer ratios} $(\ p/q/r),$ with a frequency that is a common multiple,
$\ s=CM(\ p,q,r)$. Pythagoras would like this; it is the condition for a
harmonious 3-note chord.

At $O(\gamma^{4}),$ we obtain a class of exact solutions we call $Spin(4,C)$
$Vortices$; "vortex atoms" with dense nuclei of matter currents flowing in the
$+T$ direction, outward from the big bang, and diffuse shells of returning
currents, with charges $+Z$ \ and $-Z$, respectively [M.C.5]. Kelvin would
like this.

Behind all this algebraic structure lives a simple physical picture :
\textit{each chiral pair, } $q_{I}\equiv(l\otimes_{I}r),$ \textit{acts as}
\textit{a mirror for the other 3 chiral pairs, }bootstrapping from noise a
resonant $s$- cycle.

\bigskip

\section{Reflection Varieties and Particle Masses}

The Dirac operator, $D\equiv\sigma^{\alpha}\partial_{\alpha}:e^{\beta
}\hookleftarrow q^{\beta}$, assigns a spacetime differential to an
infinitesimal displacements in the Clifford algebra [ BDS]; [G- M].

But on \textit{boundary} \textit{caustics,} $\gamma_{3-J}\subset D$ $B_{4-J} $
the spin map, $S^{\ast}=$ $\mathbf{d}\zeta=[\partial_{\alpha}\zeta^{\beta
}]\sigma_{\beta}e^{\alpha},\ $\newline becomes singular, with rank (4 -J).
Here, some steps in internal phase no longer pull back to spacetime
increments. Meanwhile, inside there are $s$ bispinor sheets for each spin
direction in the spin bundle over the particle support, $B_{4-J}$ .

For a volume element, $e^{4}$, to contribute to the action, the product of the
4 reflection operators in it must be a scalar. Physically, each cycle of 4
"interference gratings" $\ell\otimes r$ ---\emph{including} the curved
gratings involving \emph{matter spinors}, must close to form a
\emph{resonator}, with a net loop transfer function proportional to the
identity, $\sigma_{o}.$This closure constraint admits only a few sets of
integers $\{p,q,r,s\}$ characterizing the possible symmetry groups of singular
loci and isophase contours for particle wave functions: the\textit{\ Coxeter
groups, }$R_{s},$ [Coxeter] with their invariant polynomials in 4 complex
variables, the \textit{Breiskorn varieties }[Milnor] :%

\begin{equation}
R_{s}=\left\langle p,q,r\right\rangle _{s};\ \ \ \ \ \ \ \ B(x,y,z,T)\equiv
(x^{p}+y^{q}+z^{r}+T^{-s})=const.
\end{equation}

These are \emph{fibred knots.} For example, $\left\langle 2,2,3\right\rangle
_{3}\cap\mathbb{S}_{3}\left(  1\right)  $ is a trefoil knot, with its isophase
fibers, $\ln f\left(  x,y,z\right)  =const$, making three vertical twists
about the singular filament, $f\left(  x,y,z\right)  \equiv x^{2}+y^{2}%
+z^{3}=0,$ over 2 horizontal circuits.

The isophase contours over position $\ x^{\alpha}\in M$ seem to cross in the
projections,
\[
\Pi:x^{\alpha}+iy^{\alpha}\longrightarrow x^{\alpha},\text{\qquad}%
\zeta^{\alpha}=\theta^{\alpha}+i\varphi^{\alpha}\longrightarrow\theta^{\alpha
}\text{,}%
\]
from phase space to spacetime, like the crisscrossing rays in a 3D
kaleidoscope. These apparent crossings are resolved by lifting via $\Pi^{-1}
$: i.e. by separating overlapping C-algebra valued wave vectors in the
"quiver" of \ spin waves over $x^{\alpha}.$

It turns out [Cox] that the Coxeter groups varieties $\left\langle
p,q,r\right\rangle _{s}$ exhaust the topological types of \emph{resolvable
singularites}. This is just one aspect of

\textit{"the profound connections between the critical points of functions,
quivers, caustics, wavefronts, regular polyhedra,... and the \qquad\qquad
theory of groups generated by reflections"} [Arnold 2 ]. The profound
connections\textit{\ }that are important here are

\ \qquad\qquad\qquad\qquad\qquad\qquad\qquad\qquad\qquad\ \ \ \ 

\begin{enumerate}
\item $L$ or $R$ multiplication by a \textit{spacelike} C vector gives a $L$-
or $R$-\emph{helicity twist} about axis $\mathbf{\hat{\ell}}$ or
$\mathbf{\hat{r}}$ by angle $\frac{\lambda}{2}$or $\frac{\rho}{2}:$
\end{enumerate}

$\ \ \ \ \ell^{\prime}=\mathbf{L}\left(  \lambda\right)  \ell\equiv\exp\left(
\frac{i\lambda}{2}\mathbf{\hat{\ell}}\cdot\mathbf{\sigma}\right)  \ell
;$\ \ $\overline{r}^{\prime}=\bar{r}\mathbf{\bar{R}}(\rho)\equiv r\exp\left(
\frac{i\rho}{2}\mathbf{\hat{r}}\cdot\mathbf{\sigma}\right)  \Rightarrow
q^{\prime}=\left(  \ell\otimes\bar{r}\right)  ^{\prime}=L\left(  \ell
\otimes\bar{r}\right)  \bar{R}$ $.$

\qquad\qquad\qquad\ \ \ 

$\ \ \ L$ or $R$ action generated by a \textit{null }C vector gives an
additional $U(1)$ twist about the $T$ axis:

$\ \ \ \ \ \ell^{\prime}=L\ell=\exp\frac{i\lambda}{2}\left[  \pm\sigma_{0}\pm
i\mathbf{\hat{\ell}}\cdot\mathbf{\sigma}\right]  $ ,$\ \ \ \ \bar{r}^{\prime
}=\bar{r}\bar{R}=\bar{r}\exp\frac{i\rho}{2}\left[  \pm\sigma_{0}\pm
i\mathbf{\hat{r}}\cdot\mathbf{\sigma}\right]  $ $\ \Rightarrow q^{\prime
}\equiv\left(  \ell\otimes\bar{r}\right)  ^{\prime}=L\left(  \ell\otimes
\bar{r}\right)  \bar{R}$ $.$

\begin{enumerate}
\item[2.] Conjugation by a spacelike C vector reflects a flag (a $\mathbf{3}$
vector, $\mathbf{q,}$ and its normal frame) in a mirror with unit normal
$\mathbf{a:}$\ \ \ \ \ \ \ \ \ \ \ 
\end{enumerate}

$\ \ \ \ \ \ \ \ \ \mathbf{q}^{\prime}=-\mathbf{aqa}^{-1}=[i\mathbf{a]q[}%
i\mathbf{a}^{-1}]=[i\mathbf{a](}l\otimes r)\mathbf{[}i\mathbf{a}%
^{-1}]\Rightarrow l^{\prime}=[i\mathbf{a];}$ $r^{\prime}=r\mathbf{[}%
i\mathbf{a}^{-1}].$

An ordinary (period -2) reflection reverses the flag $L\leftrightarrows R,$
preserves function values, but reverses differentials, creating a
\textit{singularity} on the mirror plane. A domain $B_{3}$ \ bounded by
mirrors (like a laser cavity) becomes a \textit{resonator}: it traps waves at
its fundamental frequency or its \ harmonics to create a standing wave.\ 

\begin{enumerate}
\item[3.] Reflections in mirror planes $P_{\perp}$ and $Q_{\perp}$ that
intersect at dihedral angle $\frac{\theta}{2}$ give a \textit{rotation} by
$\ \theta$ around the (spacelike) axis. $\mathbf{a}=P_{\perp}\cap Q_{\perp}
$:$\ q^{\substack{^{\prime} \\^{\prime}}}=\ell^{\prime}\otimes\bar{r}^{\prime
}=L\left(  \ell\otimes\bar{r}\right)  L^{-1};$ $\ \ L=\exp\left(
\frac{i\lambda}{2}\mathbf{\hat{\ell}}\cdot\mathbf{\sigma}\right)  .$

\item[4.] $L$ or $R$ action by $\ $the complex Clifford ($\mathfrak{C}C$)
vector, $\exp\frac{i\theta}{p}\left[  \pm\sigma_{0}\pm i\mathbf{\hat{\ell}%
}\cdot\mathbf{\sigma}\right]  ,$ gives a \emph{period-}$p$ reflection.\ It
takes $p$ repeated reflections to close a spatial cycle; this first happens
for $\theta=\pi,$ making an\ image with \emph{dihedral} symmetry, $D_{p}$.
Parallel mirrors a distance $\frac{\bigtriangleup}{2}$ apart generate
\emph{translations} of $~\bigtriangleup$.

\item[6.] On $\ [M_{\#}]_{diag}\equiv\lbrack S_{l}(a_{\#})\times S_{3}%
(a_{\#)}]_{diag}$, multiple reflections in 3 planes that all intersect in
\emph{one point}\ form a 3D\textit{\ kaleidoscope }in spin space. Its image is
a \emph{discrete subgroup}, $R_{s}\subset U\left(  1\right)  \times SU\left(
2\right)  $, provided that the three dihedral angles,$\ \left(  \frac{\pi}%
{p},\frac{\pi}{q},\frac{\pi}{r}\right)  $ and the multiplicity, $s,$obey the
\emph{closure} constraint: : to commute, all 4 arguments above must be
multiples of $\pi$ :
\end{enumerate}

\begin{equation}%
\begin{array}
[c]{c}%
R_{P}\equiv\exp\left(  i\pi sp^{-1}P\right)  ,\quad R_{Q}\equiv\exp\left(
i\pi sq^{-1}Q\right)  ,\quad R_{R}\equiv\exp\left(  i\pi sr^{-1}R\right)
;\quad R_{S}\\
\equiv\exp\left(  i\pi sr^{-1}T\right)  ;R_{P}^{p}=R_{Q}^{q}=R_{R}^{r}%
=R_{S}^{-s}=R_{P}R_{Q}R_{R}R_{S}=-1\\
=\exp\frac{1}{2}\left(  n+1\right)  s\pi\Rightarrow s\left(  p^{-1}%
+q^{-1}+r^{-1}-s^{-1}\right)  =\left(  n+\frac{1}{2}\right)  .
\end{array}
\end{equation}

Here $\left(  p,q,r\right)  $ are integers, and $s,$ a common multiple,
$\ s=\ C.M.\left(  p,q,r\right)  :$ the \emph{multiplicity, }or
\textit{Coxeter number }of the \textit{Coxeter group.} These\textit{\ }are the
$s-$ fold covers of the Rotation, Dihedral, Tetrahedral, Octahedral, and
Icosahedra1 groups\ [Coxeter] ,%

\[
\left\langle p,q,r\right\rangle _{s}\subset\{A_{p},D_{p},T,E_{6},E_{8}\}.
\]

The cycle of $s$ reflections closes after period $t=\tau$. For the odd spin
structure [Geroch] this is the time it takes for a lightlike phase front,
$\theta=const.,$ to \ circumnavigate a closed light cone,$\ [S_{l}\times
S_{3}(a_{\#})]_{diag}$, \ \textit{twice . }Patching principle \emph{(P2) }says
that, for a periodic solution to match the vacuum on the boundary $\gamma
_{3}=\partial B_{4},$ the \ \textit{frequency,} $\omega(s,n)$ inside the
particle's world tube must be an \textit{even }harmonic of $\ $the vacuum
frequency, $(2a_{\#})^{-1}$:%

\[
\omega(s,n)=[2n(s/2)^{3}+1](2a_{\#})^{-1}=\ [n(s/2)^{3}+\frac{1}{2}%
](a_{\#})^{-1}.
\]

Each even increment, $\Delta n=2,$ adds a mass increment of
\begin{equation}
m=(s/2)^{3}(2a_{\#})^{-1}.
\end{equation}
For $s=2,$ this is the\textit{\ }mass\textit{\ }of\textit{\ }an
\textit{electron}, governed by the\textit{\ massive Dirac }equations, $(21).$
The critical radius, $a_{\#}$ $,$ is the time taken for a lightlike pulse to
traverse\ the electron's radius, in the (continuum) limit.

But isn't each cycle of Bragg reflections a discrete process? Yes. But we must
sum over all of them to get the wave equation for the matter fields; just as
we sum over random walks to get the heat equation. The "random walk"
underlying the Dirac system - the \textit{Dirac propagator} - is
the\textit{\ sum over all null zigzag histories} connecting the initial and
final states [Feynman], [Penrose l],[Ord ]. We explain, briefly here; see II
for details [M.C.3].

So far, the nonlinear 8-spinor mixing has appeared as multilinear mode-mode
coupling in \textit{spin space}. In the spacetime picture, light-like spin
rays propagate between localized scattering vertices.

Penrose calls a vertex where a L-chirality spinor reflects from a Bragg mirror
into an oppositely-propagating R- chirality one (or vica-versa).
A\textit{\ mass scattering}; is a\textit{\ null zigzag }a \textit{pair} of
mass scatterings, $L\rightarrow R\rightarrow L;$\textit{\ }the discrete
version\textit{\ }of a \textit{fold}. To close a cycle of null zigzags,
\textit{each} chiral component must return to its original value. This happens
only after a common multiple (c. m.) of the three binary reflection degrees,
$s=cm(p,q,r)$. But it takes only $\frac{s}{2}$ reflections to restore a
bispinor state; $R_{\frac{s}{2}}:\left(  \ell\oplus r\right)  \rightarrow
\left(  r\oplus\ell\right)  ,$ for $\frac{s}{2}$ odd. The energy - the
3-volume in spin-space- is counted according to its multiplicity, $s$: the
number of \ spin-space sheets above the particle's support. .

In the simplest case, the \textit{free electron}, $e^{-}\equiv(\mathbf{l}%
\oplus\mathbf{r})\in\left\langle 2,2,2\right\rangle _{2}$, all 3 dihedral
angles are $\frac{\pi}{2}$. The 3 pairs of vacuum spinors which trap the
matter pair inside a 3-cube form opposing pairs of \textit{corner-cube
reflectors. }

As we decrease one of the dihedral angles, we get a 3- cycle at $\frac{\pi}%
{3}$; 3 sheets bounded by a \textit{tuck }caustic ref. [Arnold]. But the cycle
generated by \textit{both} reflections doesn't close up again until we reach
their least-common multiple (lcm), $2\cdot3=6.$ giving a 6-fold cover of the
reflection group: the Coxeter group, $<2,2,3>_{6},$ with multiplicity $6$. We
identify this as the \textit{muon; }and the next closed reflection cycle,
$<2,3,4>_{12}$as the \textit{tauon }. More generally,

\begin{center}
\textit{\ \ A massive lepton, meson, or hadron is composed of }$\ J=1,2,$%
\textit{or }$3$\textit{\ pairs of oppositely-propagating bispinor pairs,}

\textit{\ trapped inside a timelike world tube by Bragg reflections off
interference gratings with }$(4-J)$ \textit{vacuum pairs on its boundary. }
\end{center}

What is new here is that the reflection groups $<p,q,r>_{s}$of multiplicity
\ $s=(2,3,4,5,6,12,30)$ not only \textit{classify }the elementary particles,
but give their \textit{mass ratios}, (24)%

\begin{equation}
\frac{m}{m_{e}}=(\frac{s}{2})^{3}.
\end{equation}

\textit{These agree with the observed mass ratios (table III) within a few
percent }(except for the pions-which are off by \symbol{126}25\% ).

\noindent$%
\begin{array}
[c]{c}%
\text{\textbf{Table III: Spin-}}J\text{ \textbf{Resonances:}}\\
\\
\text{Codimension-}J\text{ singularities in the }U\left(  1\right)  \times
SU\left(  2\right)  \text{ phase, };\\
\text{with wave fronts }x^{p}+y^{q}+z^{r}+T^{-s}=const\\
\text{the Brieskorn varieties. Each represents a closed cycle of Bragg
reflections of a }\\
\text{chiral pair of matter spinors, }\left(  \psi_{I},\psi^{I}\right)
,\text{off the interference gratings between }\\
\text{the remaining }\left(  J-1\right)  \text{ matter pairs and }\left(
4-J\right)  \text{ perturbed vacuum pairs.}\\%
\begin{array}
[c]{ccccccc}%
\text{Particle} &  &
\begin{array}
[c]{c}%
\text{Binary}\\
\text{Group}%
\end{array}
&  &
\begin{array}
[c]{c}%
\text{Coxeter}\\
\text{Numbers}%
\end{array}
&  & \frac{m}{m_{e}}\\
&  & H\subset\left[  SU\left(  2\right)  \right]  ^{J} &  & s:\left\langle
p,q,r\right\rangle _{s} &  &
\begin{array}
[c]{cc}%
\left(  \frac{S}{2}\right)  ^{3} & \text{obs}%
\end{array}
\\
e^{-} &  & D_{2} &  & \left\langle 2,2,2\right\rangle _{2} &  &
\begin{array}
[c]{cc}%
1 & 1
\end{array}
\\
&  & D_{p} &  & \left\langle 2,2,p\right\rangle _{p} &  & \\
&  & T &  & \left\langle 2,3,3\right\rangle _{6} &  & \\
\mu^{-} &  & O &  & \left\langle 2,3,4\right\rangle _{12} &  &
\begin{array}
[c]{cc}%
216 & 207
\end{array}
\\
\tau^{-} &  & I &  & \left\langle 2,3,5\right\rangle _{30} &  &
\begin{array}
[c]{cc}%
3375 & 3478
\end{array}
\\
\pi^{-} &  & D_{3}\otimes\bar{D}_{4} &  & \left\langle 2,2,3\right\rangle
_{3}\otimes\left\langle 2,2,4\right\rangle _{4} & d\bar{u} &
\begin{array}
[c]{cc}%
216 & 275
\end{array}
\\
k^{-} &  & D_{4}\otimes\bar{D}_{5} &  & \left\langle 2,2,5\right\rangle
_{5}\otimes\left\langle 2,2,4\right\rangle _{4} & s\bar{u} &
\begin{array}
[c]{cc}%
1000 & 975
\end{array}
\\
D_{s}^{-} &  & D_{5}\otimes\bar{D}_{6} &  & \left\langle 2,2,5\right\rangle
_{5}\otimes\left\langle 2,2,6\right\rangle _{6} & s\bar{c} &
\begin{array}
[c]{cc}%
3375 & 3647
\end{array}
\\
n_{c} &  & D_{6}\otimes\bar{D}_{6} &  & \left\langle 2,2,6\right\rangle
_{6}\otimes\left\langle 2,2,6\right\rangle _{6} & c\bar{c} &
\begin{array}
[c]{cc}%
5832 & 5686
\end{array}
\\
p^{+} &  & D_{4}\otimes D_{4}\otimes D_{3} &  & \left\langle
2,2,4\right\rangle _{4}\otimes\left\langle 2,2,3\right\rangle _{3}%
\otimes\left\langle 2,2,4\right\rangle _{4} & \left[  u,d\right]  u &
\begin{array}
[c]{cc}%
1728 & 1836
\end{array}
\end{array}
\end{array}
$

\bigskip

In the quantum calculation (III) we sum over histories in "imaginary time", T:
all possible chains of null zigzags connecting the initial and final states [MC3].

Microscopically, it seem, the whole world, both outside and inside the world
tubes of massive particles, resolves into a network of light-like spinors, and
their scattering vertices: their multilinear interactions. \ 

\ 

\section{{\protect\LARGE Conclusions and Open Question}}

Spin Principle \textbf{Pl} says that the 8-spinor bundle, $\mathbf{8}%
$\textbf{,} is the physical reality; and that the action is just its volume in
spin space. Our spacetime 4- fold, $M,$ and the particle wave functions,
$\Psi,$ are horizontal and vertical projections of a minimal-surface in spin
space: the \textit{spinfoam. }The regular stratum, or \textit{vacuum},
$D^{o},$ can be combed parallel locally by path-dependent phase differentials,
$\mathbf{d\zeta}_{I}\ =\Psi^{I}\mathbf{d\Psi}_{I},$ by spin connections: the
vector potentials. Their spin curvatures, $\Psi^{I}\mathbf{dd\Psi}_{I}$ , are
the fields, If these carry a nontrivial flux (topological charge) over the
boundary, it \textit{must }enclose a singularity-at least, in the
\textit{projection}, $\pi:$ $\mathbf{8\rightarrow}M$ : a \textit{caustic.}

These are characterized by their symmetry groups in spin space, and there are
only a few admissible types: the Coxeter groups, $\langle p,q,r\rangle_{s}$.
Their wave functions are Brieskorn varieties: fibred knots, whose isophase
contours and normal rays ("lines of force") radiate and terminate on
singularities. \ Their masses - i.e. their Noether time- translation charges,
turn out to be $m=(s/2)^{3},$ in natural units of \ $2a_{\#}{}^{-1}$; the mass
of the electron $(s=2):$%

\[
\frac{m}{m_{e}}=(s/2)^{3}.
\]

Why should the reflection groups -the same groups that
classify\textit{\ resolvable} \textit{singularities, regular polyhedra, Lie
algebras, quivers, frieze patterns, honeycombs,} \textit{crystals, and
caustics}- classify the \textit{elementary particles}? Because they all arise
from the generic structures of singularities in flows.

Like heat flow resolves into random walks, at the critical scale, $a_{\#},$
the \ $\mathbf{8-}$ spinor flow resolves into a microhistory of \textit{null
zigzags}. In each discrete history, the \textit{multiplicity}, $s$- the number
of null zig-zags it takes to close a cycle- must be a common multiple of the
reflection degrees $p,q,$and\ $r$. This results in an image in spin space like
that formed by light rays crisscrossing in a 3D kaleidoscope, with mirrors at
angles $\frac{\pi}{p},\frac{\pi}{q},\frac{\pi}{r}.$

What is subtle and beautiful about this picture is

1) how self-consistent cycles of $J$ chiral pairs of matter waves and $(4-J)$
vacuum pairs "bootstrap" each other into existence as the radius passes
through $\gamma=1,$ where $T=a_{\#};$ the critical radius for the inflationary
phase transition (\textrm{III}).

2) How the $J-$dimensional critical modes that "crystalize out" at
$O(\gamma^{J+1}),$ \textit{program the multilinear} \textit{couplings }of
modes at the next shorter scale, much as a volume hologram couples input to
output waves. This results in the ramification of patterns at smaller and
smaller scales, much like the main sequence of wavenumber-doubling
\ bifurcations leading to turbulence.

Is this what we're seeing in the sequence of $\ l=(200,400,800...)$ modes in
the Cosmic Microwave Background near the time of decoupling; or in the
foam-like structure of incident $J=(1,2,3)-$ branes in the large-scale
distribution of galaxies?

Perhaps the regular background of vacuum spinors is the\textit{\ dark energy}-
the invisible Dirac sea, on which the wave functions of visible matter ride
like waves on the surface of the ocean. Since the Dirac mass term is created
by products of vacuum spinors, these might be called \textit{dark matter}.
This picture not only shows how the "distant masses" endow particles with
their rest masses, but closely approximates the measured particle
masses\textit{.}

\end{document}